\documentclass[namedreferences]{solarphysics}
\pdfoutput=1

\usepackage[optionalrh]{spr-sola-addons} 
\usepackage[pdftex]{graphicx}        
\usepackage{color}           
\usepackage{url}             
\usepackage[breaklinks=true]{hyperref}

\newcommand{{\kms}}{    {$\mathrm {km\,s^{-1}}$}}
\newcommand{{\ms}}{    {$\mathrm {m\,s^{-1}}$}}

\begin{document}

\begin{article}

\begin{opening}

\title{The Relationship between Plasma Flow Doppler Velocities and Magnetic
Field Parameters During the Emergence of Active Regions at the Solar
Photospheric Level}

\author{A.~\surname{Khlystova}$^{1}$
       }
\runningauthor{A. Khlystova}
\runningtitle{Relationship between Velocities and Magnetic Field Parameters in
the Emerging Flux Regions}

   \institute{$^{1}$ The Institute of Solar-Terrestrial Physics, Siberian
Branch,
Russian Academy of Sciences \\
                     email: \url{hlystova@iszf.irk.ru}
             }

\begin{abstract}
A statistical study has been carried out of the relationship between plasma flow
Doppler velocities and magnetic field parameters during the emergence
of active regions at the solar photospheric level with data acquired by
the \textit{Michelson Doppler Imager} (MDI) onboard the \textit{Solar and
Heliospheric Observatory} (SOHO). We have investigated 224 emerging active
regions with different spatial scales and positions on the solar disc.
The following relationships for the first hours of the emergence of active
regions have been analysed: i) of peak negative Doppler velocities with
the position of the emerging active regions on the solar disc; ii) of peak
plasma upflow and downflow Doppler velocities with the magnetic flux growth
rate and magnetic field strength for the active regions emerging near
the solar disc centre (the vertical component of plasma flows); iii) of
peak positive and negative Doppler velocities with the magnetic flux
growth rate and magnetic field strength for the active regions emerging near the
limb (the horizontal component of plasma flows); iv) of the magnetic flux
growth rate with the density of emerging magnetic flux; v) of the
Doppler velocities and magnetic field parameters for the first hours of
the appearance of active regions with the total unsigned magnetic flux
at the maximum of their development. 
\end{abstract} 

\end{opening}

\section{Introduction}
     \label{sec:intro}

\par Solar magnetic flux emerges at various spatial scales ({\it e.g.}
\opencite{par09}). \inlinecite{gol98} revealed a power-law relation between the
maximum magnetic flux and the lifetime of active regions. \inlinecite{ots11}
found a relation between the flux growth rate and the maximum magnetic flux in
the form of a power-law function.

\par Measurements of Doppler velocities indicated that there is a vertical
plasma upflow up to 1\,{\kms} at the polarity inversion line when
active regions emerge in the solar photosphere
\cite{bra85a,bra85b,tar89,lit98,str99,kub03,gug06,gri07,gri09}.
\inlinecite{gri07} revealed high Doppler velocities of about
1.7\,{\kms} at the beginning of the emergence of the powerful active region NOAA
10488 at heliographic coordinates N08 E31 (B$_{0}+$4.9$^{\circ}$). The
observed Doppler velocities have not been compared with the parameters
of the magnetic fields.

\par The cold and dense plasma contained inside magnetic flux
emerging into the solar atmosphere flows down along the magnetic field lines
with velocities of up to 2\,{\kms} at the photospheric level
\cite{gop67,gop69,kaw76,bac78,zwa85,bra85a,bra85b,bra85c,lit98,sol03,lag07,xu10}. 
Observations of pores revealed a statistical relation between the
Doppler velocity of plasma downflow and the magnetic field strength.
\inlinecite{bon91}, studying a large pore, found that the strong magnetic
fields correlate linearly with downflow Doppler velocities in
the form B\,=\,500\,V, where B is in Gauss and V is in {\kms}.
\inlinecite{kei99} found that downflow Doppler velocities inside pores
correlate positively with the magnetic field strength. \inlinecite{cho10}
analysed small pores less than 2$''$, which were not related to 
magnetic flux emergence, by using high spatial resolution data
from \textit{Hinode}. The authors revealed a negative correlation
between positive Doppler velocities and the magnetic field strength,
{\it i.e.} small downflow Doppler velocities corresponded to strong
magnetic fields. \inlinecite{gri11} have shown that in the emerging
active region during the formation of pores the plasma downflow
Doppler velocity increases linearly with the magnetic field
strength.

\par Estimations of horizontal velocities in the emerging active regions were
performed indirectly. A wide range of velocity values from 0.1 to 5.5\,{\kms}
were obtained by analysing displacements of magnetic field structures
\cite{fra72,sch73,har73,cho87,bar90,str99,hag01,gri09,ots11}. Considering the
separation of opposite polarity poles in 24 bipolar pairs \inlinecite{cho87}
found no relation of horizontal velocities neither with the average magnetic
field strength nor with the total unsigned magnetic flux.
\inlinecite{ots11} studied 101 emerging flux regions with various spatial
scales. They showed a power-law relation with negative index between the
horizontal velocities of the separation of opposite polarity poles and
the maximum value of magnetic flux.

\par \inlinecite{khl11} presented a statistical study the photospheric
Doppler velocities in 83 active regions with a total unsigned
magnetic flux above $10^{21}$\,Mx. A centre to limb dependence
of negative Doppler velocities was established, which indicates that the
horizontal velocities of plasma outflows are higher than the vertical
upflow velocities during the first hours of the emergence of active
regions.

\par In this article we study the statistical relations between Doppler
velocities and magnetic field parameters during the emergence of active regions.

\section{Data Analysis}
     \label{sec:data}

\par We used full-disc SOHO/MDI data \cite{sch95}. The temporal resolution
of the photospheric magnetograms and Dopplergrams is 1 minute, while that of
the continuum is 96 minutes. The spatial resolution of the data is
4$''$. Magnetograms with a 1.8.2 calibration level were used \cite{ulr09}.

\par We have cropped a region of emerging magnetic flux from
a time sequence of data taking into account its displacement caused by
solar rotation. This displacement was calculated by the differential
rotation law for photospheric magnetic fields \cite{sno83} and the
cross-correlation analysis of two magnetograms adjacent in time. A
precise spatial superposition of the data was achieved by cropping fragments
with identical coordinates from simultaneously acquired magnetograms and
Dopplergrams. For correct processing of the data, we chose the cropped region in
a way that it excluded the area outside the limb. For the Dopplergrams the
contribution of solar differential rotation and other factors distorting the
Doppler velocity signal were removed by using the technique described
in \inlinecite{gri07}.

\par The sequence of the fragments obtained has been used to calculate the temporal
variations of the parameters under study. The calculation area was limited to
the region of the emerging magnetic flux. The boundary of emerging flux
region was visually inspected.

\par The total unsigned magnetic flux of active regions was calculated
inside isolines $\pm$60\,G taking into account the projection effect
and assuming that the magnetic field vector is perpendicular to the solar surface:
\begin{eqnarray}
\Phi = S_{0} \sum_{i=1}^N \frac{|B_{i}|}{cos \theta_{i}},\
\end{eqnarray}
where $\Phi$ is the total unsigned magnetic flux in Mx, $S_{0}$ is the
area of the solar surface of pixel in the centre of solar disc in $cm^{2}$, $N$
is the number of pixels where $|B_{i}|$ $>$ 60\,G, $B_{i}$ is the line of sight
magnetic field strength of $ith$ pixel in $G$, $\theta_{i}$ is the heliocentric
angle of $ith$ pixel. The maximum total unsigned magnetic flux
$\Phi_{max}$ was determined by the maximum inflection point in the
increase of the magnetic flux curve or by the last value of the datasets (for
active regions passing beyond the west limb or with insufficiently
downloaded data). The signal background that existed before the emergence of
active regions was subtracted from the maximum total unsigned magnetic
flux. Magnetic saturation in measurements of the magnetic field
strength in SOHO/MDI was controlled \cite{liu07}; it was attained only
in three of the active regions considered.

\par The total unsigned magnetic flux growth rate {\it d$\Phi$/dt} and
absolute value of the maximum magnetic field strength $H_{max}$ were calculated
in the first 12 hours of the emergence of the magnetic flux for large and small
active regions or from beginning to maximum of the total unsigned
magnetic flux for ephemeral active regions.

\par Photospheric plasma flows are characterised by the peak (absolute 
maximum) values of negative and positive Doppler velocities $V_{max-}$ and
$V_{max+}$ which were determined in the first 12 hours of the magnetic flux
emergence for large and small active regions or from beginning to maximum of the
total unsigned magnetic flux for ephemeral active regions. The magnetic
flux emergence begins with the appearance of the loop apex where the magnetic
field is horizontal; therefore, the time interval we analysed started
30 minutes before the appearance of the line of sight magnetic fields.
Continuum images were used to control the possible formation of 
sunspots with Evershed flows in the first hours of the emergence of active
regions to eliminate their contribution. The maximum Doppler velocity
of these flows in SOHO/MDI data with low spatial resolution can reach
2\,{\kms} \cite{bai98}.

\par The position of the active region on the solar disc is expressed by the
heliocentric angle $\theta$:
\begin{eqnarray}
\theta = \arcsin(r/R),
\end{eqnarray}
where $r$ is the distance from the disc centre to the place of the
active region emergence and $R$ is the solar radius. $\theta$ is also
approximately the angle between the normal to the surface and
the line of sight component of the emerging magnetic flux.

\par We performed a regression analysis of the data. There is at least one
inflection point in the considered dependencies. Therefore, linear or quadratic
polynomials were used as approximation functions. The choice of the
polynomial degree was based on the minimum values of the standard
deviations. The results are given in the form of polynomial equations
which, according to the F-statistic, are significant. The confidence intervals for
the means were calculated with a confidence probability of 99\%.

\section{The Investigated Active Regions}

\par
\begin{figure}
\centerline{
\includegraphics[width=\textwidth]{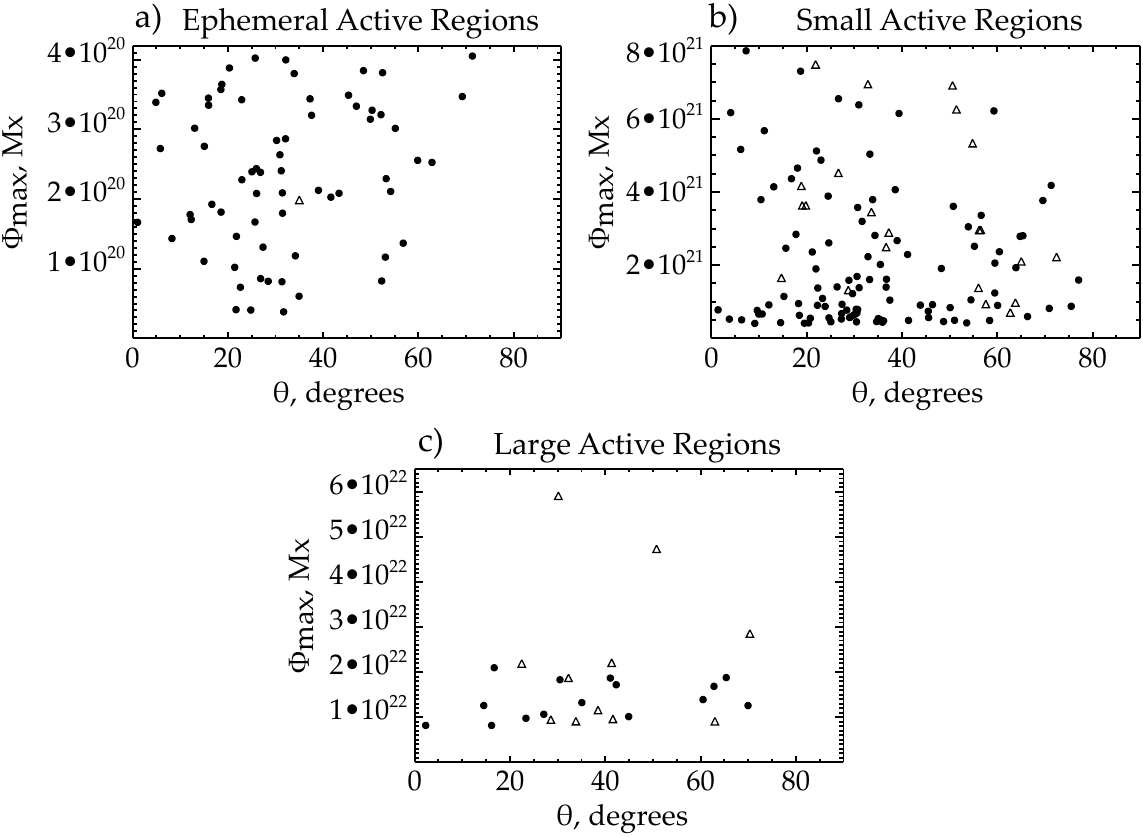}
}
\caption{
Maximum total unsigned magnetic flux of the (a) ephemeral, (b)
small and (c) large active regions versus the heliocentric angle that
corresponds to the beginning of the emergence of magnetic fluxes. The
total magnetic flux maximum was determined by the maximum inflection
point in the increase of the magnetic flux curve (marked by points) or
by the last measurement (marked by triangles). Active regions with magnetic
saturation are also marked by triangles. In active regions marked by triangles,
the maximum of the total unsigned magnetic flux may be higher.
        }
\label{fig:flux}
\end{figure}

 We have studied 224 active regions which emerged on the visible side of the
solar disc from 1999 to 2008 (see Figure~\ref{fig:flux}). The selected active
regions have different spatial scales and emerge at different distances from the
solar disc centre. They are isolated from high concentrations of
existing magnetic fields (the presence of single poles with total
unsigned magnetic flux not higher than $0.5\times10^{21}$\,Mx in the
region of the direct emergence of the active region for the first hours was
allowed). Sufficiently complete data series with a temporal resolution of 1
minute for the first hours of magnetic flux emergence are available for the
selected objects.

The statistical analysis includes the following.

\begin{enumerate}
 \item According to the maximum value of the total unsigned magnetic flux (in the two polarities):
\begin{itemize}
 \item 68 ephemeral active regions ($2.6\times10^{18}$\,Mx $ < \Phi_{max} <
4.07\times10^{20}$\,Mx);
 \item 130 small active regions ($4.07\times10^{20}$\,Mx $ < \Phi_{max} <
8\times10^{21}$\,Mx);
 \item 26 large active regions ($\Phi_{max} > 8\times10^{21}$\,Mx).
\end{itemize}

\par The limits of the total unsigned magnetic flux for ephemeral active regions
were taken from \inlinecite{hag01}. The limit between the large and small active
regions is considered to be the maximum magnetic flux in one polarity; it is taken equal to
$5\times10^{21}$\,Mx \cite{gar84,zwa87}. In the present investigation the limit
between the large and small active regions was chosen at the total unsigned
magnetic flux in two polarities, equal to $8\times10^{21}$\,Mx, since all active
regions with the total unsigned magnetic flux above this level contain sunspots
and their spatial size exceeds the size of a supergranule.

 \item According to the distances from the solar disc centre:
\begin{itemize}
 \item 72 active regions near the solar disc centre ($\theta <
25^{\circ}$);
 \item 98 active regions at medium distance from disc centre ($25^{\circ} <
\theta < 50^{\circ}$);
 \item 54 active regions near the limb ($\theta > 50^{\circ}$).
\end{itemize}
\end{enumerate}

\par
\begin{figure}
\centerline{
\includegraphics[height=4.5cm]{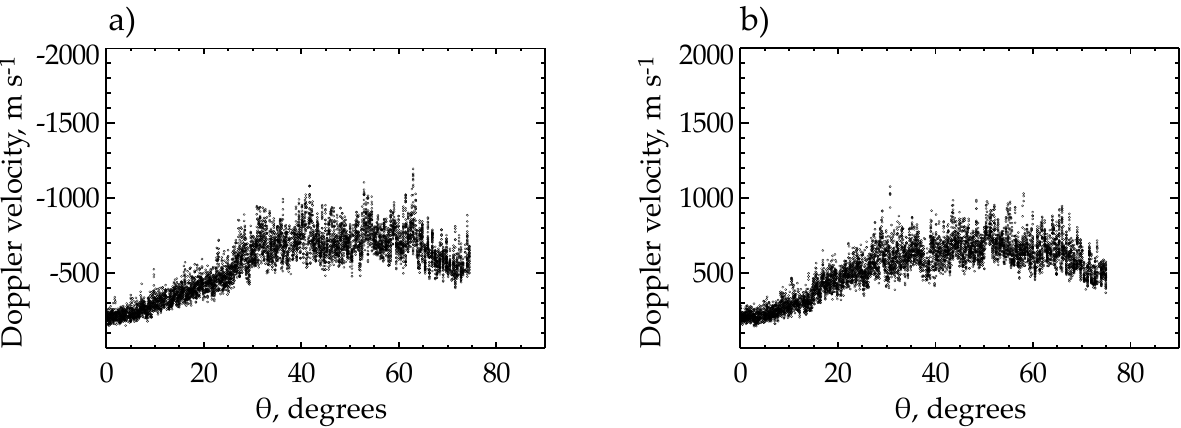}
}
\caption{
Peak (a) negative and (b) positive Doppler velocities of the
convective flows in the quiet Sun versus the heliocentric angle for SOHO/MDI
data.
        }
\label{fig:quiet-sun}
\end{figure}

 The maximum Doppler velocities of the convection flow in the quiet
Sun were calculated at different distances from the disc centre using
SOHO/MDI data (Figure~\ref{fig:quiet-sun}). Dopplergrams with 1 minute temporal
resolution for the period 18\,--\,23 April 2001 were used. The peak
values of negative and positive Doppler velocities were measured
within a region of size 40$''\times$40$''$ whose displacement was tracked based
on the differential rotation law of Doppler structures in the photosphere
\cite{sno90}. The selected region was going through the W00\,--\,W75 range of
longitudes along the equator during the period under study. The contribution of
solar differential rotation and other factors distorting the Doppler
velocity signal were removed by using the technique described by
\inlinecite{gri07}.

\par In Figure~\ref{fig:quiet-sun} one observes that Doppler velocities of
convection flows increases with the heliocentric angle. For the points with
$\theta > 60^{\circ}$ there is a tendency for Doppler velocities to decrease.
This may be due to the Doppler velocities being measured in higher atmospheric
layers near the limb, due to the increase in the optical thickness, where the
atmosphere is convectively stable and convection flows are damped.

\section{Results}

\subsection{The Centre to Limb Dependence of Negative Doppler Velocities}

\par
\begin{figure}
\centerline{
\includegraphics[width=\textwidth]{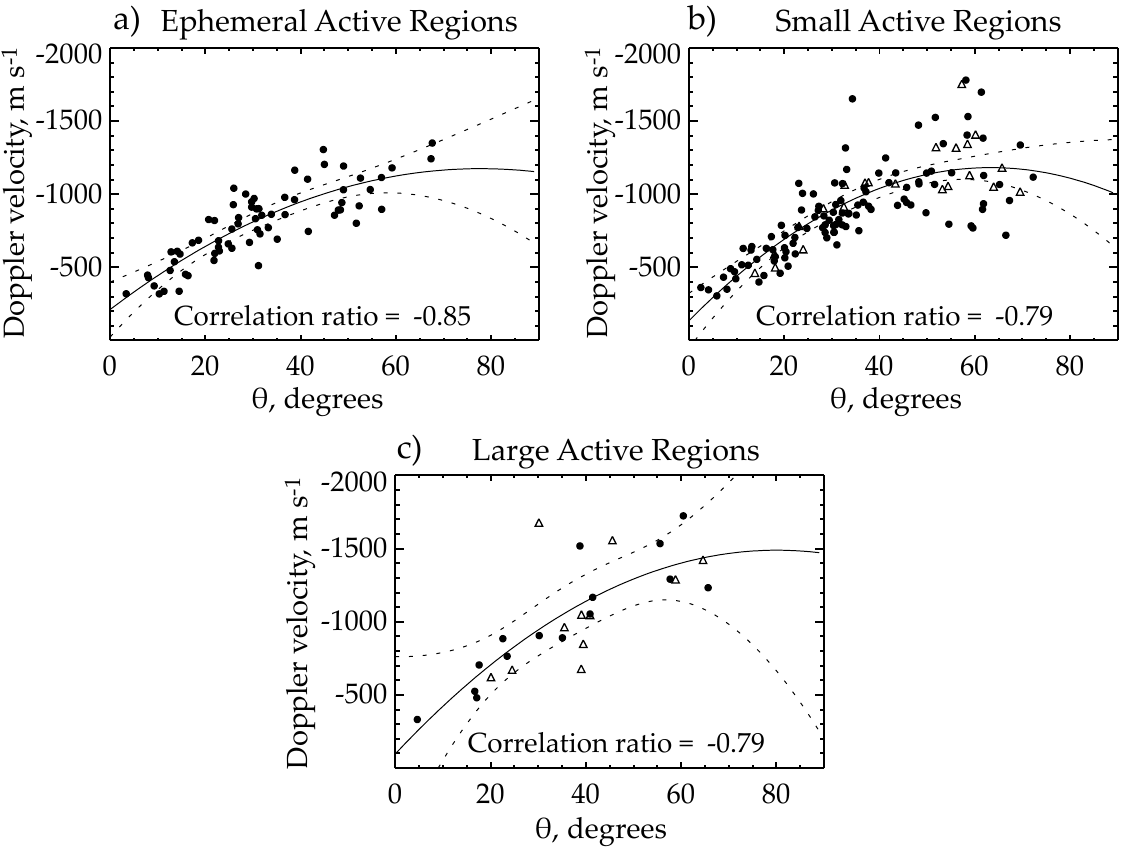}
}
\caption{
Peak negative Doppler velocities (plasma motion towards the
observer) in the first hours of the emergence of (a) ephemeral, (b) small
and (c) large active regions versus the heliocentric angle that corresponds to
the position of the region at this time. The solid line corresponds
to Equations (\ref{eqn:centre-limb-ephemeral}) in plot (a),
(\ref{eqn:centre-limb-small}) in plot (b) and (\ref{eqn:centre-limb-large}) in
plot (c); dotted lines correspond to 99\% confidence intervals for the means.
        }
\label{fig:vmin-theta}
\end{figure}

 The peak values of negative Doppler velocities (plasma
motion toward the observer) in the first hours of magnetic flux
emergence of ephemeral, small and large active regions show an increase with
the heliocentric angle (Figure~\ref{fig:vmin-theta}). The regression analysis of
the data yielded Equations (\ref{eqn:centre-limb-ephemeral}) for
ephemeral, (\ref{eqn:centre-limb-small}) for small, and
(\ref{eqn:centre-limb-large}) for large active regions:
\begin{eqnarray}
V_{max-} = -211.32 - 24.77 \theta + 0.16 \theta^{2},
\label{eqn:centre-limb-ephemeral}
\end{eqnarray}
\begin{eqnarray}
V_{max-} = -134.93 - 33.05 \theta + 0.26 \theta^{2},
\label{eqn:centre-limb-small}
\end{eqnarray}
\begin{eqnarray}
V_{max-} = -96.35 - 34.75 \theta + 0.22 \theta^{2},
\label{eqn:centre-limb-large}
\end{eqnarray}
where $V_{max-}$ is the peak negative Doppler velocity in {\ms}
and $\theta$ is the heliocentric angle in degrees that
corresponds to the position of the region at $V_{max-}$. Correlation ratios
show a strong relation between the considered parameters: $-$0.85 for
ephemeral, $-$0.79 for small and $-$0.79 for large active regions. The
dependencies obtained indicate that during the emergence of active regions the
horizontal velocities of plasma outflows ($\theta > 50^{\circ}$) exceed the
vertical upflow velocities ($\theta < 25^{\circ}$). Generally, the
Doppler velocities accompanying the emergence of active regions are
higher than those of convective flows, however, sometimes they are
comparable. For small and large active regions the horizontal
Doppler velocity component substantially exceeds horizontal
Doppler velocities of the convection flows in the quiet Sun,
which are less than 1200\,{\ms} (Figure~\ref{fig:vmin-theta}\,b,\,c, and
Figure~\ref{fig:quiet-sun}\,a the points with $\theta > 50^{\circ}$).

\subsection{Vertical Doppler Velocities}

\par
\begin{figure}
\centerline{
\includegraphics[height=4.5cm]{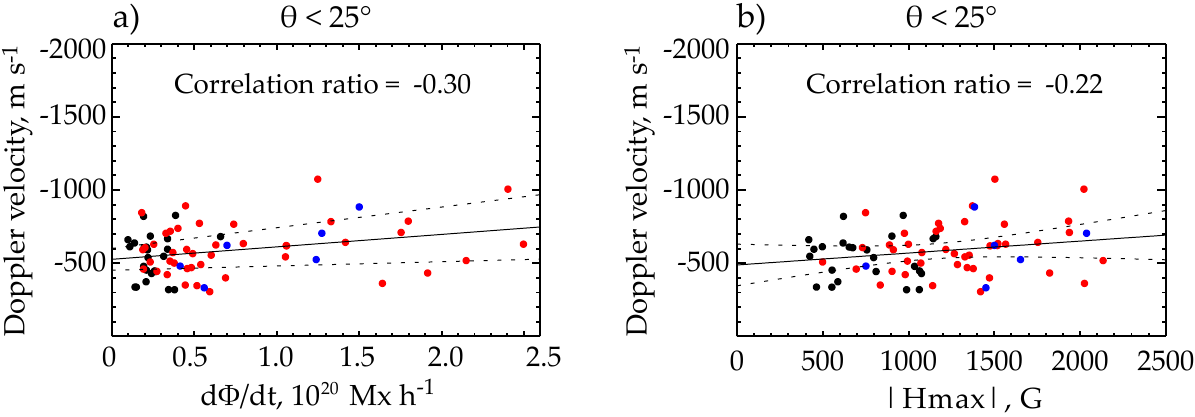}
}
\caption{
Peak negative Doppler velocities (plasma motion toward the
observer) versus (a) the magnetic flux growth rate and (b) the maximum
magnetic field strength in the first hours of the emergence of active
regions near the solar disc centre ($\theta < 25^{\circ}$). Black
symbols are for ephemeral active regions, red symbols for small active
regions and blue symbols for large active regions.
        }
\label{fig:vmin}
\end{figure}

 For the active regions emerging near the solar disc centre
($\theta < 25^{\circ}$), peak negative Doppler velocities
(plasma upflow) have a high dispersion and correlate neither with the
magnetic flux growth rate (the correlation ratio is $-$0.30) nor with
the maximum magnetic field strength (the correlation ratio is $-$0.22)
(Figure~\ref{fig:vmin}). The dependencies obtained imply that the upflow
Doppler velocities do not characterise the power of the emerging magnetic
flux.

\begin{figure}
\centerline{
\includegraphics[height=4.5cm]{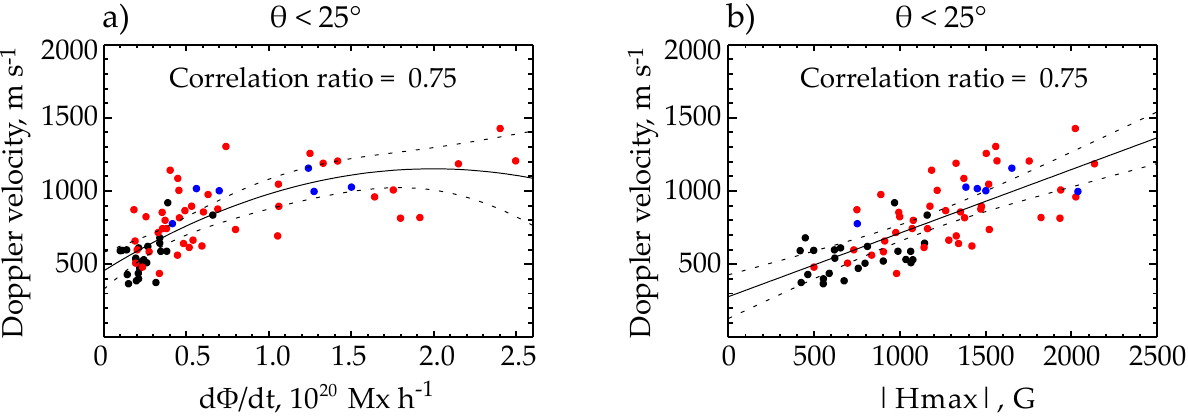}
}
\caption{
Peak positive Doppler velocities (plasma motion away from the
observer) versus (a) the magnetic flux growth rate and (b) the maximum
magnetic field strength in the first hours of the emergence of active
regions near the solar disc centre ($\theta < 25^{\circ}$). The
solid line corresponds to Equations (\ref{eqn:vmax}) in plot (a) and
(\ref{eqn:vmax_h}) in plot (b); dotted lines correspond to 99\% confidence
intervals for the means. The symbol colours are the same as in
Figure~\ref{fig:vmin}.
        }
\label{fig:vmax}
\end{figure}

\par For the active regions emerging in the central part of the solar disc
($\theta < 25^{\circ}$), the peak positive Doppler velocities
(plasma downflow) are related quadratically to the magnetic
flux growth rate (the correlation ratio is 0.75) and linearly with the
maximum magnetic field strength (the correlation ratio is 0.75)
(Figure~\ref{fig:vmax}). The linear dependence shown on Figure~\ref{fig:vmax}\,b
confirms the results obtained previously by \inlinecite{bon91},
\inlinecite{kei99}, \inlinecite{gri11}. The Doppler velocity values
substantially exceed convection Doppler velocities in the
quiet Sun (Figure~\ref{fig:vmax} and Figure~\ref{fig:quiet-sun}\,b the points
with $\theta < 25^{\circ}$).

\par Regression analysis of the data yielded the following equations:
\begin{eqnarray}
V_{max+} = 455.82 + 6.98\times10^{-18}d\Phi/dt -
1.75\times10^{-38}(d\Phi/dt)^{2},
\label{eqn:vmax}
\end{eqnarray}
\begin{eqnarray}
V_{max+} = 277.94 + 0.43 \times H_{max},
\label{eqn:vmax_h}
\end{eqnarray}
where $V_{max+}$ is the peak positive Doppler velocity in
{\ms}, $d\Phi/dt$ is the magnetic flux growth rate in {$\mathrm
{Mx\,h^{-1}}$} and $H_{max}$ is the maximum magnetic field strength in G.

\par The draining of the cold plasma, being carried out by the emerging
magnetic flux from the convective zone into the solar atmosphere, is
considered to be the main reason for the plasma observed downflows.

\begin{figure}
\centerline{
\includegraphics[height=4.5cm]{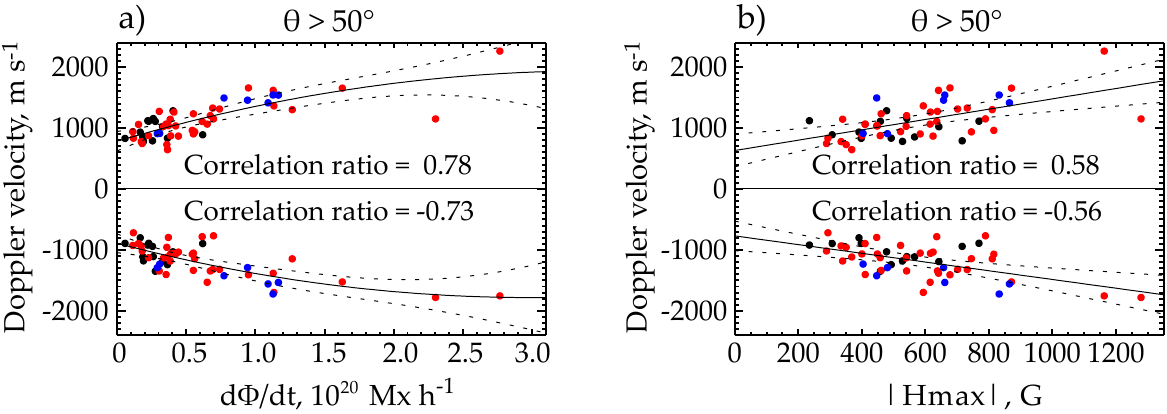}
}
\caption{
The peak positive and negative Doppler velocities versus (a) the
magnetic flux growth rate and (b) maximum magnetic field strength in the first
hours of the emergence of active regions near the limb ($\theta >
50^{\circ}$). The solid line corresponds to Equations
(\ref{eqn:vhor-max}) and (\ref{eqn:vhor-min}) in plot (a) and Equations
(\ref{eqn:vhor-max_h}) and (\ref{eqn:vhor-min_h}) in plot (b), dotted lines
correspond to 99\% confidence intervals for the means. The symbol colours are
the same as in Figure~\ref{fig:vmin}.
        }
\label{fig:vhor}
\end{figure}

\subsection{Horizontal Dopper Velocity}

\par \inlinecite{khl12} has carried out a detailed analysis of photospheric
flows in four active regions emerging near the solar limb. It has been found
that extended regions of Doppler velocities with different signs are formed in
the first hours of the magnetic flux emergence in the horizontal velocity field.
The observed Doppler velocities substantially exceed the separation velocities
of the photospheric magnetic flux outer boundaries and most likely are caused by
the significant component of the velocities of plasma downflow being carried out
into the solar atmosphere by emerging magnetic flux.

\par One observes in Figure~\ref{fig:vmin-theta} that the horizontal
Doppler velocity component ($\theta > 50^{\circ}$) in the region of
emerging magnetic fluxes has a wide range of values. For the active
regions emerging near the limb ($\theta > 50^{\circ}$), the peak
positive and negative Doppler velocities of photospheric plasma flows show
a quadratic relation with the magnetic flux growth rate (correlation
ratios 0.78 and $-$0.73) and a linear relation with the maximum
magnetic field strength (correlation ratios 0.58 and $-$0.56)
(Figure~\ref{fig:vhor}). Strong magnetic fields are oriented mainly
perpendicular to the surface after the emergence. Therefore, the
magnetic field strength in active regions near the limb is 2\,--\,3 times
smaller than that one in active regions near the solar disc
centre, because of the projection effects of the magnetic field vector
to the line of sight (Figure~\ref{fig:vhor}\,b and Figure~\ref{fig:vmax}\,b).

\par Regression analysis of the data yielded the following equations:
\begin{eqnarray}
V_{max+} = 782.27 + 6.68\times10^{-18}d\Phi/dt -
9.65\times10^{-39}(d\Phi/dt)^{2},
\label{eqn:vhor-max}
\end{eqnarray}
\begin{eqnarray}
V_{max-} = -890.85 - 5.95\times10^{-18}d\Phi/dt +
9.94\times10^{-39}(d\Phi/dt)^{2},
\label{eqn:vhor-min}
\end{eqnarray}
\begin{eqnarray}
V_{max+} = 630.53 + 0.85 \times H_{max},
\label{eqn:vhor-max_h}
\end{eqnarray}
\begin{eqnarray}
V_{max-} = -770.35 - 0.71 \times H_{max},
\label{eqn:vhor-min_h}
\end{eqnarray}
where $V_{max+}$ and $V_{max-}$ are, respectively, the peak values
of positive and negative Doppler velocities in {\ms}, $d\Phi/dt$ is
the magnetic flux growth rate in {$\mathrm {Mx\,h^{-1}}$} and $H_{max}$ is
the maximum magnetic field strength in G.

\par One sees in Figure~\ref{fig:vmax}\,a and Figure~\ref{fig:vhor}\,a that
the horizontal Doppler velocities exceed the vertical downflow Doppler
velocities by a factor of 1.5\,--\,2 in the active regions with the same
magnetic flux growth rate. This supports theoretical models in which the
velocities of draining plasma have a significant horizontal component at
the beginning of the magnetic flux emergence (see \inlinecite{shi90},
\inlinecite{arch04}, \inlinecite{tor10}, \inlinecite{tor11} and others).

\begin{figure}
\centerline{
\includegraphics[height=4.5cm]{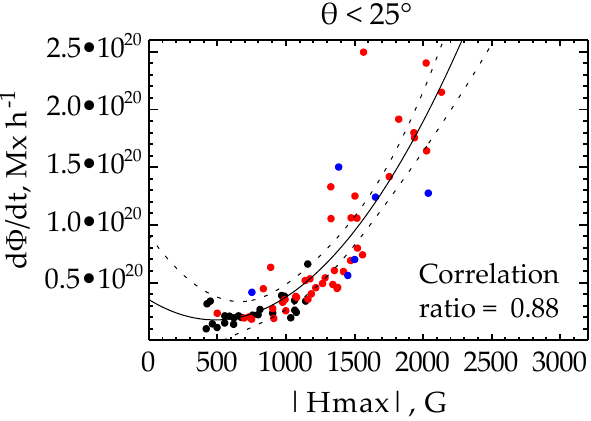}
}
\caption{
Magnetic flux growth rate versus the magnetic field strength in the first hours
of the emergence of active regions near the solar disc centre
($\theta < 25^{\circ}$). The solid line corresponds to Equation
(\ref{eqn:flux-rate}); dotted lines correspond to 99\% confidence intervals for
the means. The symbol colours are the same as in Figure~\ref{fig:vmin}.
        }
\label{fig:flux-rate}
\end{figure}

\subsection{The Relation between the Magnetic Flux Growth Rate and the Maximum
Strength of Magnetic Fields}

\par A quadratic relation with relatively low dispersion and a high
correlation ratio of 0.88 between the magnetic flux growth rate and the density
of the magnetic fields for the active regions emerging near the solar
disc centre ($\theta < 25^{\circ}$) has been found
(Figure~\ref{fig:flux-rate}). The equation derived by the regression analysis
has the following form:
\begin{eqnarray}
d\Phi/dt = 3.52\times10^{19} - 7.27\times10^{16} H_{max} + 7.49\times10^{13}
H_{max}^{2},
\label{eqn:flux-rate}
\end{eqnarray}
where $d\Phi/dt$ is the magnetic flux growth rate in the first hours of the
emergence in {$\mathrm {Mx\,h^{-1}}$} and $H_{max}$ is the maximum
magnetic field strength in G.

\par The dependence is theoretically expectable. The magnetic flux emerge
due to the action of the magnetic buoyant force which is proportional
to the square of the magnetic field density $B^{2}$
(\opencite{par55}). The higher the magnetic field strength, the
higher the magnetic pressure and the lower the gas pressure inside the magnetic
structure. The low gas pressure leads to a large magnetic buoyancy
force and, thus, to an increase in the magnetic flux
growth rate.

\subsection{The Relation of Parameters between the Beginning and Maximum
Development of Active Regions}

\begin{figure}
\centerline{
\includegraphics[width=\textwidth]{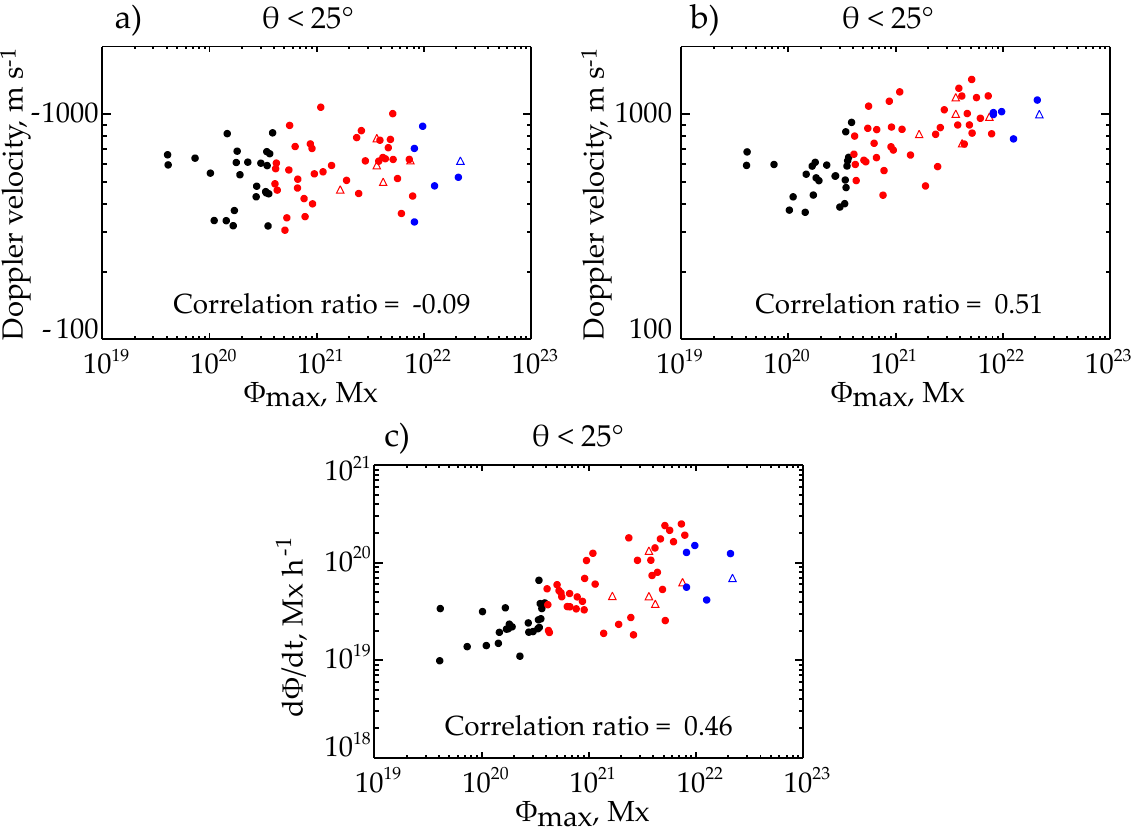}
}
\caption{
The relation between (a) peak negative Doppler velocities, (b)
peak positive Doppler velocities and (c) the magnetic flux
growth rate in the first hours of the emergence of active regions with total
unsigned magnetic flux at the maximum of their development
$\Phi_{max}$. The symbol colours are the same as in Figure~\ref{fig:vmin}. The
$\Phi_{max}$ was determined by the maximum inflection point in the
increase of the flux curve (marked by points) or by the last measurement (marked
by triangles). In active regions marked by triangles, the $\Phi_{max}$ may be
higher.
	}
\label{fig:flux-max}
\end{figure}

\par Active regions emerging near the solar disc centre ($\theta <
25^{\circ}$) are considered. Peak negative Doppler velocities in the
first hours of the magnetic flux emergence do not show any relation with
the total unsigned magnetic flux at maximum development of the
active regions (Figure~\ref{fig:flux-max}\,a). Comparison of the peak
positive Doppler velocities and the magnetic flux growth rate with the maximum
value of total unsigned magnetic flux reveals a power-law
relation with high dispersion (Figure~\ref{fig:flux-max}\,b,\,c). A
power-law relation, but with less dispersion, between the flux growth rate and
the the maximum value of magnetic flux has been found earlier
by \inlinecite{ots11}. In the present investigation the weak dependence
between the parameters under consideration can be explained by the fact
that active region magnetic flux emerges in separate fragments
on time scales from several hours to 5\,--\,7 days. Therefore, the
parameters of magnetic fields and Doppler velocities in the first hours
of emergence do not characterise an active region as a whole.

\section{Conclusions} 
  \label{S-Conclusion}

The study of plasma flow Doppler velocities and magnetic field
parameters during the emergence of active regions at the photospheric level has
revealed the following statistical relationships:

\begin{enumerate}
\item Peak values of negative Doppler velocities accompanying the
magnetic flux emergence in the first hours show an increase with the
heliocentric angle~$\theta$. The observed relation shows that horizontal
velocities of plasma outflows are higher than the vertical upflow velocities.

\item The upflow plasma Doppler velocities have no relation with
parameters of magnetic fields and, thus, do not
characterise the power of the emerging magnetic flux.

\item The downflow plasma Doppler velocities have high values,
which substantially exceed Doppler velocities of convection flows in the
quiet Sun. They are related quadratically with the magnetic flux growth rate
and linearly with the maximum magnetic field strength. The main reason for
the observed downflows is considered to be a draining of the cold
plasma, being carried out into the solar atmosphere by the emerging
magnetic flux.

\item The horizontal plasma outflow Doppler velocities are connected
quadratically with the magnetic flux growth rate and linearly with the maximum
magnetic field strength. Possibly the horizontal flows are connected with
draining of plasma which has a significant horizontal
component at the beginning of the magnetic flux emergence.

\item There is a quadratic dependence with relatively low dispersion and
a high correlation ratio of 0.88 between the magnetic flux growth rate
and the maximum strength of magnetic fields. The observed relation is consistent
with the magnetic buoyancy law.

\item There is a weak power-law dependence of peak positive Doppler
velocities and the magnetic flux growth rate in the first hours of the
emergence of active regions with the maximum value of the total
unsigned magnetic flux.

\end{enumerate}

\begin{acks}
The author is grateful to the referee for helpful comments that improved the
manuscript. The author thanks V.\,M.\,Grigor'ev, L.\,V.\,Ermakova and V.\,G.\,
Fainsh- tein for useful discussions. This work used data obtained by the SOHO/MDI
instrument. SOHO is a mission of international cooperation between ESA and NASA.
The MDI is a project of the Stanford-Lockheed Institute for
Space Research. This study was supported by RFBR grants 10-02-00607-a,
10-02-00960-a, 11-02-00333-a, 12-02-00170-a, the state contracts of the Ministry
of Education and Science of the Russian Federation No. 02.740.11.0576,
16.518.11.7065, the program of the Division of Physical Sciences of the Russian
Academy of Sciences No. 16, the Integration Project of SB RAS No. 13 and the
program of Presidium of Russian Academy of Sciences No. 22.
\end{acks}

\bibliographystyle{spr-mp-sola}
\bibliography{khlystova}

\begin{thebibliography}{46}
\ifx \bisbn   \undefined \def \bisbn  #1{ISBN #1}\fi
\ifx \binits  \undefined \def \binits#1{#1}\fi
\ifx \bauthor  \undefined \def \bauthor#1{#1}\fi
\ifx \batitle  \undefined \def \batitle#1{#1}\fi
\ifx \bjtitle  \undefined \def \bjtitle#1{\textit{#1}}\fi
\ifx \bvolume  \undefined \def \bvolume#1{\textbf{#1}}\fi
\ifx \byear  \undefined \def \byear#1{#1}\fi
\ifx \bissue  \undefined \def \bissue#1{#1}\fi
\ifx \bfpage  \undefined \def \bfpage#1{#1}\fi
\ifx \blpage  \undefined \def \blpage #1{#1}\fi
\ifx \burl  \undefined \def \burl#1{\textsf{#1}}\fi
\ifx \href  \undefined \def \href#1#2{\textsf{#2}}\fi
\ifx \doiurl  \undefined \def
  \doiurl#1{\href{http://dx.doi.org/#1}{\textsf{#1}}}\fi
\ifx \betal  \undefined \def \betal{\textit{et al.}}\fi
\ifx \binstitute  \undefined \def \binstitute#1{#1}\fi
\ifx \bctitle  \undefined \def \bctitle#1{#1}\fi
\ifx \beditor  \undefined \def \beditor#1{#1}\fi
\ifx \bpublisher  \undefined \def \bpublisher#1{#1}\fi
\ifx \bbtitle  \undefined \def \bbtitle#1{\textit{#1}}\fi
\ifx \bedition  \undefined \def \bedition#1{#1}\fi
\ifx \bseriesno  \undefined \def \bseriesno#1{\textbf{#1}}\fi
\ifx \blocation  \undefined \def \blocation#1{#1}\fi
\ifx \bsertitle  \undefined \def \bsertitle#1{\textit{#1}}\fi
\ifx \bsnm \undefined \def \bsnm#1{#1}\fi
\ifx \bsuffix \undefined \def \bsuffix#1{#1}\fi
\ifx \bparticle \undefined \def \bparticle#1{#1}\fi
\ifx \barticle \undefined \def \barticle#1{}\fi
\ifx \botherref \undefined \def \botherref#1{}\fi
\ifx \url \undefined \def \url#1{\textsf{#1}}\fi
\ifx \bchapter \undefined \def \bchapter#1{}\fi
\ifx \bbook \undefined \def \bbook#1{}\fi
\ifx \bcomment \undefined \def \bcomment#1{#1}\fi
\ifx \oauthor \undefined \def \oauthor#1{#1}\fi
\ifx \citeauthoryear \undefined \def \citeauthoryear#1{#1}\fi
\def \endbibitem {}
\ifx \bconflocation  \undefined \def \bconflocation#1{#1} \fi

\bibitem[\protect\citeauthoryear{{Archontis} \textit{et~al.}}{2004}]{arch04}
\begin{barticle}
\bauthor{\bsnm{{Archontis}}, \binits{V.}},
\bauthor{\bsnm{{Moreno-Insertis}}, \binits{F.}},
\bauthor{\bsnm{{Galsgaard}}, \binits{K.}},
\bauthor{\bsnm{{Hood}}, \binits{A.}},
\bauthor{\bsnm{{O'Shea}}, \binits{E.}}:
\byear{2004},
\batitle{{Emergence of magnetic flux from the convection zone into the
  corona}}.
\bjtitle{\aap}
\bvolume{426},
\bfpage{1047}\,--\,\blpage{1063}.
doi:\doiurl{10.1051/0004-6361:20035934}.
\end{barticle}
\endbibitem

\bibitem[\protect\citeauthoryear{{Bachmann}}{1978}]{bac78}
\begin{barticle}
\bauthor{\bsnm{{Bachmann}}, \binits{G.}}:
\byear{1978},
\batitle{{On the evolution of magnetic and velocity fields of an originating
  sunspot group}}.
\bjtitle{Bulletin of the Astronomical Institutes of Czechoslovakia}
\bvolume{29},
\bfpage{180}\,--\,\blpage{184}.
\end{barticle}
\endbibitem

\bibitem[\protect\citeauthoryear{{Bai}, {Scherrer}, and {Bogart}}{1998}]{bai98}
\begin{bchapter}
\bauthor{\bsnm{{Bai}}, \binits{T.}},
\bauthor{\bsnm{{Scherrer}}, \binits{P.H.}},
\bauthor{\bsnm{{Bogart}}, \binits{R.S.}}:
\byear{1998},
\bctitle{{The Evershed Effect: an MDI Investigation}}.
In: \beditor{\bsnm{{S.~Korzennik}}} (ed.)
\bbtitle{Structure and Dynamics of the Interior of the Sun and Sun-like Stars},
\bsertitle{ESA Special Publication}
\bseriesno{418},
\bfpage{607}\,--\,\blpage{610}.
\end{bchapter}
\endbibitem

\bibitem[\protect\citeauthoryear{{Barth} and {Livi}}{1990}]{bar90}
\begin{barticle}
\bauthor{\bsnm{{Barth}}, \binits{C.S.}},
\bauthor{\bsnm{{Livi}}, \binits{S.H.B.}}:
\byear{1990},
\batitle{{Magnetic Bipoles in Emerging Flux Regions on the Sun}}.
\bjtitle{Rev. Mex. Astron. Astrofis.}
\bvolume{21},
\bfpage{549}\,--\,\blpage{551}.
\end{barticle}
\endbibitem

\bibitem[\protect\citeauthoryear{{Bonaccini} \textit{et~al.}}{1991}]{bon91}
\begin{bchapter}
\bauthor{\bsnm{{Bonaccini}}, \binits{D.}},
\bauthor{\bsnm{{Landi Degl'Innocenti}}, \binits{E.}},
\bauthor{\bsnm{{Smaldone}}, \binits{L.A.}},
\bauthor{\bsnm{{Tamblyn}}, \binits{P.}}:
\byear{1991},
\bctitle{{High resolution spectropolarimetry of an active region.}}
In: \beditor{\bsnm{{L.~J.~November}}} (ed.)
\bbtitle{Solar Polarimetry},
\bfpage{251}\,--\,\blpage{256}.
\end{bchapter}
\endbibitem

\bibitem[\protect\citeauthoryear{{Brants}}{1985a}]{bra85a}
\begin{barticle}
\bauthor{\bsnm{{Brants}}, \binits{J.J.}}:
\byear{1985}a,
\batitle{{High-resolution spectroscopy of active regions. II Line-profile
  interpretation, applied to an emerging flux region}}.
\bjtitle{\solphys}
\bvolume{95},
\bfpage{15}\,--\,\blpage{36}.
doi:\doiurl{10.1007/BF00162633}.
\end{barticle}
\endbibitem

\bibitem[\protect\citeauthoryear{{Brants}}{1985b}]{bra85b}
\begin{barticle}
\bauthor{\bsnm{{Brants}}, \binits{J.J.}}:
\byear{1985}b,
\batitle{{High-resolution spectroscopy of active regions. III - Relations
  between the intensity, velocity, and magnetic structure in an emerging flux
  region}}.
\bjtitle{\solphys}
\bvolume{98},
\bfpage{197}\,--\,\blpage{217}.
doi:\doiurl{10.1007/BF00152456}.
\end{barticle}
\endbibitem

\bibitem[\protect\citeauthoryear{{Brants} and {Steenbeek}}{1985}]{bra85c}
\begin{barticle}
\bauthor{\bsnm{{Brants}}, \binits{J.J.}},
\bauthor{\bsnm{{Steenbeek}}, \binits{J.C.M.}}:
\byear{1985},
\batitle{{Morphological evolution of an emerging flux region}}.
\bjtitle{\solphys}
\bvolume{96},
\bfpage{229}\,--\,\blpage{252}.
doi:\doiurl{10.1007/BF00149682}.
\end{barticle}
\endbibitem

\bibitem[\protect\citeauthoryear{{Cho} \textit{et~al.}}{2010}]{cho10}
\begin{barticle}
\bauthor{\bsnm{{Cho}}, \binits{K.-S.}},
\bauthor{\bsnm{{Bong}}, \binits{S.-C.}},
\bauthor{\bsnm{{Chae}}, \binits{J.}},
\bauthor{\bsnm{{Kim}}, \binits{Y.-H.}},
\bauthor{\bsnm{{Park}}, \binits{Y.-D.}}:
\byear{2010},
\batitle{{Tiny Pores Observed by Hinode/Solar Optical Telescope}}.
\bjtitle{\apj}
\bvolume{723},
\bfpage{440}\,--\,\blpage{448}.
doi:\doiurl{10.1088/0004-637X/723/1/440}.
\end{barticle}
\endbibitem

\bibitem[\protect\citeauthoryear{{Chou} and {Wang}}{1987}]{cho87}
\begin{barticle}
\bauthor{\bsnm{{Chou}}, \binits{D.}},
\bauthor{\bsnm{{Wang}}, \binits{H.}}:
\byear{1987},
\batitle{{The separation velocity of emerging magnetic flux}}.
\bjtitle{\solphys}
\bvolume{110},
\bfpage{81}\,--\,\blpage{99}.
doi:\doiurl{10.1007/BF00148204}.
\end{barticle}
\endbibitem

\bibitem[\protect\citeauthoryear{{Frazier}}{1972}]{fra72}
\begin{barticle}
\bauthor{\bsnm{{Frazier}}, \binits{E.N.}}:
\byear{1972},
\batitle{{The Magnetic Structure of Arch Filament Systems}}.
\bjtitle{\solphys}
\bvolume{26},
\bfpage{130}\,--\,\blpage{141}.
doi:\doiurl{10.1007/BF00155113}.
\end{barticle}
\endbibitem

\bibitem[\protect\citeauthoryear{{Garcia de La Rosa}}{1984}]{gar84}
\begin{barticle}
\bauthor{\bsnm{{Garcia de La Rosa}}, \binits{J.I.}}:
\byear{1984},
\batitle{{The observation of intrinsically different emergences for large and
  small active regions}}.
\bjtitle{\solphys}
\bvolume{92},
\bfpage{161}\,--\,\blpage{172}.
doi:\doiurl{10.1007/BF00157243}.
\end{barticle}
\endbibitem

\bibitem[\protect\citeauthoryear{{Golovko}}{1998}]{gol98}
\begin{barticle}
\bauthor{\bsnm{{Golovko}}, \binits{A.A.}}:
\byear{1998},
\batitle{{Relationship between the maximum magnetic fluxes and lifetimes of
  solar active regions}}.
\bjtitle{Astron. Rep.}
\bvolume{42},
\bfpage{546}\,--\,\blpage{552}.
\end{barticle}
\endbibitem

\bibitem[\protect\citeauthoryear{{Gopasyuk}}{1967}]{gop67}
\begin{barticle}
\bauthor{\bsnm{{Gopasyuk}}, \binits{S.I.}}:
\byear{1967},
\batitle{{The velocity field in an active region at spot appearance stag.}}
\bjtitle{Izvestiya Ordena Trudovogo Krasnogo Znameni Krymskoj Astrofizicheskoj
  Observatorii}
\bvolume{37},
\bfpage{29}\,--\,\blpage{43}.
\end{barticle}
\endbibitem

\bibitem[\protect\citeauthoryear{{Gopasyuk}}{1969}]{gop69}
\begin{barticle}
\bauthor{\bsnm{{Gopasyuk}}, \binits{S.I.}}:
\byear{1969},
\batitle{{The velocity field on the two levels in the active region of July
  1966.}}
\bjtitle{Izvestiya Ordena Trudovogo Krasnogo Znameni Krymskoj Astrofizicheskoj
  Observatorii}
\bvolume{40},
\bfpage{111}\,--\,\blpage{126}.
\end{barticle}
\endbibitem

\bibitem[\protect\citeauthoryear{{Grigor'ev}, {Ermakova}, and
  {Khlystova}}{2007}]{gri07}
\begin{barticle}
\bauthor{\bsnm{{Grigor'ev}}, \binits{V.M.}},
\bauthor{\bsnm{{Ermakova}}, \binits{L.V.}},
\bauthor{\bsnm{{Khlystova}}, \binits{A.I.}}:
\byear{2007},
\batitle{{Dynamics of line-of-sight velocities and magnetic field in the solar
  photosphere during the formation of the large active region NOAA 10488}}.
\bjtitle{Astronomy Letters}
\bvolume{33},
\bfpage{766}\,--\,\blpage{770}.
doi:\doiurl{10.1134/S1063773707110072}.
\end{barticle}
\endbibitem

\bibitem[\protect\citeauthoryear{{Grigor'ev}, {Ermakova}, and
  {Khlystova}}{2009}]{gri09}
\begin{barticle}
\bauthor{\bsnm{{Grigor'ev}}, \binits{V.M.}},
\bauthor{\bsnm{{Ermakova}}, \binits{L.V.}},
\bauthor{\bsnm{{Khlystova}}, \binits{A.I.}}:
\byear{2009},
\batitle{{Emergence of magnetic flux at the solar surface and the origin of
  active regions}}.
\bjtitle{Astron. Rep.}
\bvolume{53},
\bfpage{869}\,--\,\blpage{878}.
doi:\doiurl{10.1134/S1063772909090108}.
\end{barticle}
\endbibitem

\bibitem[\protect\citeauthoryear{{Grigor'ev}, {Ermakova}, and
  {Khlystova}}{2011}]{gri11}
\begin{barticle}
\bauthor{\bsnm{{Grigor'ev}}, \binits{V.M.}},
\bauthor{\bsnm{{Ermakova}}, \binits{L.V.}},
\bauthor{\bsnm{{Khlystova}}, \binits{A.I.}}:
\byear{2011},
\batitle{{The dynamics of photospheric line-of-sight velocities in emerging
  active regions}}.
\bjtitle{Astron. Rep.}
\bvolume{55},
\bfpage{163}\,--\,\blpage{173}.
doi:\doiurl{10.1134/S1063772911020041}.
\end{barticle}
\endbibitem

\bibitem[\protect\citeauthoryear{{Guglielmino} \textit{et~al.}}{2006}]{gug06}
\begin{barticle}
\bauthor{\bsnm{{Guglielmino}}, \binits{S.L.}},
\bauthor{\bsnm{{Mart{\'{\i}}nez Pillet}}, \binits{V.}},
\bauthor{\bsnm{{Ruiz Cobo}}, \binits{B.}},
\bauthor{\bsnm{{Zuccarello}}, \binits{F.}},
\bauthor{\bsnm{{Lites}}, \binits{B.W.}}:
\byear{2006},
\batitle{{A Detailed Analysis of an Ephemeral Region .}}
\bjtitle{Memorie della Societa Astronomica Italiana Supplementi}
\bvolume{9},
\bfpage{103}\,--\,\blpage{105}.
\end{barticle}
\endbibitem

\bibitem[\protect\citeauthoryear{{Hagenaar}}{2001}]{hag01}
\begin{barticle}
\bauthor{\bsnm{{Hagenaar}}, \binits{H.J.}}:
\byear{2001},
\batitle{{Ephemeral Regions on a Sequence of Full-Disk Michelson Doppler Imager
  Magnetograms}}.
\bjtitle{\apj}
\bvolume{555},
\bfpage{448}\,--\,\blpage{461}.
doi:\doiurl{10.1086/321448}.
\end{barticle}
\endbibitem

\bibitem[\protect\citeauthoryear{{Harvey} and {Martin}}{1973}]{har73}
\begin{barticle}
\bauthor{\bsnm{{Harvey}}, \binits{K.L.}},
\bauthor{\bsnm{{Martin}}, \binits{S.F.}}:
\byear{1973},
\batitle{{Ephemeral Active Regions}}.
\bjtitle{\solphys}
\bvolume{32},
\bfpage{389}\,--\,\blpage{402}.
doi:\doiurl{10.1007/BF00154951}.
\end{barticle}
\endbibitem

\bibitem[\protect\citeauthoryear{{Kawaguchi} and {Kitai}}{1976}]{kaw76}
\begin{barticle}
\bauthor{\bsnm{{Kawaguchi}}, \binits{I.}},
\bauthor{\bsnm{{Kitai}}, \binits{R.}}:
\byear{1976},
\batitle{{The velocity field associated with the birth of sunspots}}.
\bjtitle{\solphys}
\bvolume{46},
\bfpage{125}\,--\,\blpage{135}.
doi:\doiurl{10.1007/BF00157559}.
\end{barticle}
\endbibitem

\bibitem[\protect\citeauthoryear{{Keil} \textit{et~al.}}{1999}]{kei99}
\begin{barticle}
\bauthor{\bsnm{{Keil}}, \binits{S.L.}},
\bauthor{\bsnm{{Balasubramaniam}}, \binits{K.S.}},
\bauthor{\bsnm{{Smaldone}}, \binits{L.A.}},
\bauthor{\bsnm{{Reger}}, \binits{B.}}:
\byear{1999},
\batitle{{Velocities in Solar Pores}}.
\bjtitle{\apj}
\bvolume{510},
\bfpage{422}\,--\,\blpage{443}.
doi:\doiurl{10.1086/306549}.
\end{barticle}
\endbibitem

\bibitem[\protect\citeauthoryear{{Khlystova}}{2011}]{khl11}
\begin{barticle}
\bauthor{\bsnm{{Khlystova}}, \binits{A.}}:
\byear{2011},
\batitle{{Center-limb dependence of photospheric velocities in regions of
  emerging magnetic fields on the Sun}}.
\bjtitle{\aap}
\bvolume{528},
\bfpage{A7}.
doi:\doiurl{10.1051/0004-6361/201015765}.
\end{barticle}
\endbibitem

\bibitem[\protect\citeauthoryear{{Khlystova}}{2012}]{khl12}
\begin{botherref}
\oauthor{\bsnm{{Khlystova}}, \binits{A.}}:
2012,
{The Horizontal Component of Photospheric Plasma Flows During the Emergence of
  Active Regions on the Sun}.
\textit{\solphys, in Topical Issue ``Advances of European Solar Physics''}.
doi:\doiurl{10.1007/s11207-012-0181-8}.
\end{botherref}
\endbibitem

\bibitem[\protect\citeauthoryear{{Kubo}, {Shimizu}, and {Lites}}{2003}]{kub03}
\begin{barticle}
\bauthor{\bsnm{{Kubo}}, \binits{M.}},
\bauthor{\bsnm{{Shimizu}}, \binits{T.}},
\bauthor{\bsnm{{Lites}}, \binits{B.W.}}:
\byear{2003},
\batitle{{The Evolution of Vector Magnetic Fields in an Emerging Flux Region}}.
\bjtitle{\apj}
\bvolume{595},
\bfpage{465}\,--\,\blpage{482}.
doi:\doiurl{10.1086/377333}.
\end{barticle}
\endbibitem

\bibitem[\protect\citeauthoryear{{Lagg} \textit{et~al.}}{2007}]{lag07}
\begin{barticle}
\bauthor{\bsnm{{Lagg}}, \binits{A.}},
\bauthor{\bsnm{{Woch}}, \binits{J.}},
\bauthor{\bsnm{{Solanki}}, \binits{S.K.}},
\bauthor{\bsnm{{Krupp}}, \binits{N.}}:
\byear{2007},
\batitle{{Supersonic downflows in the vicinity of a growing pore. Evidence of
  unresolved magnetic fine structure at chromospheric heights}}.
\bjtitle{\aap}
\bvolume{462},
\bfpage{1147}\,--\,\blpage{1155}.
doi:\doiurl{10.1051/0004-6361:20054700}.
\end{barticle}
\endbibitem

\bibitem[\protect\citeauthoryear{{Lites}, {Skumanich}, and {Martinez
  Pillet}}{1998}]{lit98}
\begin{barticle}
\bauthor{\bsnm{{Lites}}, \binits{B.W.}},
\bauthor{\bsnm{{Skumanich}}, \binits{A.}},
\bauthor{\bsnm{{Martinez Pillet}}, \binits{V.}}:
\byear{1998},
\batitle{{Vector magnetic fields of emerging solar flux. I. Properties at the
  site of emergence}}.
\bjtitle{\aap}
\bvolume{333},
\bfpage{1053}\,--\,\blpage{1068}.
\end{barticle}
\endbibitem

\bibitem[\protect\citeauthoryear{{Liu}, {Norton}, and {Scherrer}}{2007}]{liu07}
\begin{barticle}
\bauthor{\bsnm{{Liu}}, \binits{Y.}},
\bauthor{\bsnm{{Norton}}, \binits{A.A.}},
\bauthor{\bsnm{{Scherrer}}, \binits{P.H.}}:
\byear{2007},
\batitle{{A Note on Saturation Seen in the MDI/SOHO Magnetograms}}.
\bjtitle{\solphys}
\bvolume{241},
\bfpage{185}\,--\,\blpage{193}.
doi:\doiurl{10.1007/s11207-007-0296-5}.
\end{barticle}
\endbibitem

\bibitem[\protect\citeauthoryear{{Otsuji} \textit{et~al.}}{2011}]{ots11}
\begin{barticle}
\bauthor{\bsnm{{Otsuji}}, \binits{K.}},
\bauthor{\bsnm{{Kitai}}, \binits{R.}},
\bauthor{\bsnm{{Ichimoto}}, \binits{K.}},
\bauthor{\bsnm{{Shibata}}, \binits{K.}}:
\byear{2011},
\batitle{{Statistical Study on the Nature of Solar-Flux Emergence}}.
\bjtitle{\pasj}
\bvolume{63},
\bfpage{1047}\,--\,\blpage{1057}.
\end{barticle}
\endbibitem

\bibitem[\protect\citeauthoryear{{Parker}}{1955}]{par55}
\begin{barticle}
\bauthor{\bsnm{{Parker}}, \binits{E.N.}}:
\byear{1955},
\batitle{{The Formation of Sunspots from the Solar Toroidal Field.}}
\bjtitle{\apj}
\bvolume{121},
\bfpage{491}\,--\,\blpage{507}.
doi:\doiurl{10.1086/146010}.
\end{barticle}
\endbibitem

\bibitem[\protect\citeauthoryear{{Parnell} \textit{et~al.}}{2009}]{par09}
\begin{barticle}
\bauthor{\bsnm{{Parnell}}, \binits{C.E.}},
\bauthor{\bsnm{{DeForest}}, \binits{C.E.}},
\bauthor{\bsnm{{Hagenaar}}, \binits{H.J.}},
\bauthor{\bsnm{{Johnston}}, \binits{B.A.}},
\bauthor{\bsnm{{Lamb}}, \binits{D.A.}},
\bauthor{\bsnm{{Welsch}}, \binits{B.T.}}:
\byear{2009},
\batitle{{A Power-Law Distribution of Solar Magnetic Fields Over More Than Five
  Decades in Flux}}.
\bjtitle{\apj}
\bvolume{698},
\bfpage{75}\,--\,\blpage{82}.
doi:\doiurl{10.1088/0004-637X/698/1/75}.
\end{barticle}
\endbibitem

\bibitem[\protect\citeauthoryear{{Scherrer} \textit{et~al.}}{1995}]{sch95}
\begin{barticle}
\bauthor{\bsnm{{Scherrer}}, \binits{P.H.}},
\bauthor{\bsnm{{Bogart}}, \binits{R.S.}},
\bauthor{\bsnm{{Bush}}, \binits{R.I.}},
\bauthor{\bsnm{{Hoeksema}}, \binits{J.T.}},
\bauthor{\bsnm{{Kosovichev}}, \binits{A.G.}},
\bauthor{\bsnm{{Schou}}, \binits{J.}},
\bauthor{\bsnm{{Rosenberg}}, \binits{W.}},
\bauthor{\bsnm{{Springer}}, \binits{L.}},
\bauthor{\bsnm{{Tarbell}}, \binits{T.D.}},
\bauthor{\bsnm{{Title}}, \binits{A.}},
\bauthor{\bsnm{{Wolfson}}, \binits{C.J.}},
\bauthor{\bsnm{{Zayer}}, \binits{I.}},
\bauthor{\bsnm{{MDI Engineering Team}}}:
\byear{1995},
\batitle{{The Solar Oscillations Investigation - Michelson Doppler Imager}}.
\bjtitle{\solphys}
\bvolume{162},
\bfpage{129}\,--\,\blpage{188}.
doi:\doiurl{10.1007/BF00733429}.
\end{barticle}
\endbibitem

\bibitem[\protect\citeauthoryear{{Schoolman}}{1973}]{sch73}
\begin{barticle}
\bauthor{\bsnm{{Schoolman}}, \binits{S.A.}}:
\byear{1973},
\batitle{{Videomagnetograph Studies of Solar Magnetic Fields. II: Field Changes
  in an Active Region}}.
\bjtitle{\solphys}
\bvolume{32},
\bfpage{379}\,--\,\blpage{388}.
doi:\doiurl{10.1007/BF00154950}.
\end{barticle}
\endbibitem

\bibitem[\protect\citeauthoryear{{Shibata} \textit{et~al.}}{1990}]{shi90}
\begin{barticle}
\bauthor{\bsnm{{Shibata}}, \binits{K.}},
\bauthor{\bsnm{{Nozawa}}, \binits{S.}},
\bauthor{\bsnm{{Matsumoto}}, \binits{R.}},
\bauthor{\bsnm{{Sterling}}, \binits{A.C.}},
\bauthor{\bsnm{{Tajima}}, \binits{T.}}:
\byear{1990},
\batitle{{Emergence of solar magnetic flux from the convection zone into the
  photosphere and chromosphere}}.
\bjtitle{\apjl}
\bvolume{351},
\bfpage{L25}\,--\,\blpage{L28}.
doi:\doiurl{10.1086/185671}.
\end{barticle}
\endbibitem

\bibitem[\protect\citeauthoryear{{Snodgrass}}{1983}]{sno83}
\begin{barticle}
\bauthor{\bsnm{{Snodgrass}}, \binits{H.B.}}:
\byear{1983},
\batitle{{Magnetic rotation of the solar photosphere}}.
\bjtitle{\apj}
\bvolume{270},
\bfpage{288}\,--\,\blpage{299}.
doi:\doiurl{10.1086/161121}.
\end{barticle}
\endbibitem

\bibitem[\protect\citeauthoryear{{Snodgrass} and {Ulrich}}{1990}]{sno90}
\begin{barticle}
\bauthor{\bsnm{{Snodgrass}}, \binits{H.B.}},
\bauthor{\bsnm{{Ulrich}}, \binits{R.K.}}:
\byear{1990},
\batitle{{Rotation of Doppler features in the solar photosphere}}.
\bjtitle{\apj}
\bvolume{351},
\bfpage{309}\,--\,\blpage{316}.
doi:\doiurl{10.1086/168467}.
\end{barticle}
\endbibitem

\bibitem[\protect\citeauthoryear{{Solanki} \textit{et~al.}}{2003}]{sol03}
\begin{barticle}
\bauthor{\bsnm{{Solanki}}, \binits{S.K.}},
\bauthor{\bsnm{{Lagg}}, \binits{A.}},
\bauthor{\bsnm{{Woch}}, \binits{J.}},
\bauthor{\bsnm{{Krupp}}, \binits{N.}},
\bauthor{\bsnm{{Collados}}, \binits{M.}}:
\byear{2003},
\batitle{{Three-dimensional magnetic field topology in a region of solar
  coronal heating}}.
\bjtitle{\nat}
\bvolume{425},
\bfpage{692}\,--\,\blpage{695}.
doi:\doiurl{10.1038/nature02035}.
\end{barticle}
\endbibitem

\bibitem[\protect\citeauthoryear{{Strous} and {Zwaan}}{1999}]{str99}
\begin{barticle}
\bauthor{\bsnm{{Strous}}, \binits{L.H.}},
\bauthor{\bsnm{{Zwaan}}, \binits{C.}}:
\byear{1999},
\batitle{{Phenomena in an Emerging Active Region. II. Properties of the Dynamic
  Small-Scale Structure}}.
\bjtitle{\apj}
\bvolume{527},
\bfpage{435}\,--\,\blpage{444}.
doi:\doiurl{10.1086/308071}.
\end{barticle}
\endbibitem

\bibitem[\protect\citeauthoryear{{Tarbell} \textit{et~al.}}{1989}]{tar89}
\begin{bchapter}
\bauthor{\bsnm{{Tarbell}}, \binits{T.D.}},
\bauthor{\bsnm{{Topka}}, \binits{K.}},
\bauthor{\bsnm{{Ferguson}}, \binits{S.}},
\bauthor{\bsnm{{Frank}}, \binits{Z.}},
\bauthor{\bsnm{{Title}}, \binits{A.M.}}:
\byear{1989},
\bctitle{{High - resolution observations of emerging magnetic flux}}.
In: \beditor{\bsnm{{O.~von der Luehe}}} (ed.)
\bbtitle{High spatial resolution solar observations},
\bfpage{506}\,--\,\blpage{520}.
\end{bchapter}
\endbibitem

\bibitem[\protect\citeauthoryear{{Toriumi} and {Yokoyama}}{2010}]{tor10}
\begin{barticle}
\bauthor{\bsnm{{Toriumi}}, \binits{S.}},
\bauthor{\bsnm{{Yokoyama}}, \binits{T.}}:
\byear{2010},
\batitle{{Two-step Emergence of the Magnetic Flux Sheet from the Solar
  Convection Zone}}.
\bjtitle{\apj}
\bvolume{714},
\bfpage{505}\,--\,\blpage{516}.
doi:\doiurl{10.1088/0004-637X/714/1/505}.
\end{barticle}
\endbibitem

\bibitem[\protect\citeauthoryear{{Toriumi} \textit{et~al.}}{2011}]{tor11}
\begin{barticle}
\bauthor{\bsnm{{Toriumi}}, \binits{S.}},
\bauthor{\bsnm{{Miyagoshi}}, \binits{T.}},
\bauthor{\bsnm{{Yokohama}}, \binits{T.}},
\bauthor{\bsnm{{Isobe}}, \binits{H.}},
\bauthor{\bsnm{{Shibata}}, \binits{K.}}:
\byear{2011},
\batitle{{Dependence of the Magnetic Energy of Solar Active Regions on the
  Twist Intensity of the Initial Flux Tubes}}.
\bjtitle{\pasj}
\bvolume{63},
\bfpage{407}\,--\,\blpage{415}.
\end{barticle}
\endbibitem

\bibitem[\protect\citeauthoryear{{Ulrich} \textit{et~al.}}{2009}]{ulr09}
\begin{barticle}
\bauthor{\bsnm{{Ulrich}}, \binits{R.K.}},
\bauthor{\bsnm{{Bertello}}, \binits{L.}},
\bauthor{\bsnm{{Boyden}}, \binits{J.E.}},
\bauthor{\bsnm{{Webster}}, \binits{L.}}:
\byear{2009},
\batitle{{Interpretation of Solar Magnetic Field Strength Observations}}.
\bjtitle{\solphys}
\bvolume{255},
\bfpage{53}\,--\,\blpage{78}.
doi:\doiurl{10.1007/s11207-008-9302-9}.
\end{barticle}
\endbibitem

\bibitem[\protect\citeauthoryear{{Xu}, {Lagg}, and {Solanki}}{2010}]{xu10}
\begin{barticle}
\bauthor{\bsnm{{Xu}}, \binits{Z.}},
\bauthor{\bsnm{{Lagg}}, \binits{A.}},
\bauthor{\bsnm{{Solanki}}, \binits{S.K.}}:
\byear{2010},
\batitle{{Magnetic structures of an emerging flux region in the solar
  photosphere and chromosphere}}.
\bjtitle{\aap}
\bvolume{520},
\bfpage{A77}.
doi:\doiurl{10.1051/0004-6361/200913227}.
\end{barticle}
\endbibitem

\bibitem[\protect\citeauthoryear{{Zwaan}}{1987}]{zwa87}
\begin{barticle}
\bauthor{\bsnm{{Zwaan}}, \binits{C.}}:
\byear{1987},
\batitle{{Elements and patterns in the solar magnetic field}}.
\bjtitle{Annual review of astronomy and astrophysics}
\bvolume{25},
\bfpage{83}\,--\,\blpage{111}.
doi:\doiurl{10.1146/annurev.aa.25.090187.000503}.
\end{barticle}
\endbibitem

\bibitem[\protect\citeauthoryear{{Zwaan}, {Brants}, and {Cram}}{1985}]{zwa85}
\begin{barticle}
\bauthor{\bsnm{{Zwaan}}, \binits{C.}},
\bauthor{\bsnm{{Brants}}, \binits{J.J.}},
\bauthor{\bsnm{{Cram}}, \binits{L.E.}}:
\byear{1985},
\batitle{{High-resolution spectroscopy of active regions. I - Observing
  procedures}}.
\bjtitle{\solphys}
\bvolume{95},
\bfpage{3}\,--\,\blpage{14}.
doi:\doiurl{10.1007/BF00162632}.
\end{barticle}
\endbibitem

\end{thebibliography}

\end{article}
\end{document}